\documentclass[conference]{IEEEtran}
\IEEEoverridecommandlockouts

\usepackage{cite}
\usepackage{amsmath,amssymb,amsfonts}
\usepackage{algorithmic}
\usepackage{graphicx}
\usepackage{textcomp}
\usepackage{xcolor}
\usepackage{booktabs}
\usepackage{url}
\usepackage{xurl}

\def\BibTeX{{\rm B\kern-.05em{\sc i\kern-.025em b}\kern-.08em
    T\kern-.1667em\lower.7ex\hbox{E}\kern-.125emX}}
\begin{document}

\title{Group Dynamics of Engagement with AI Topics on Bluesky\\
\thanks{This work was supported in part by the U.S. National Science Foundation Division of Mathematical Sciences under Grant No. 2042413 and by the Air Force Office of Scientific Research Multidisciplinary University Research Initiative under Grant No. FA9550-22-1-0380.}
}

\author{
\IEEEauthorblockN{1\textsuperscript{st} Bradley Huynh}
\IEEEauthorblockA{\textit{Dept. of Electrical, Computer \& Energy Eng.} \\
\textit{University of Colorado Boulder}\\
Boulder, CO, USA\\
ORCID: 0009-0004-3087-9738}
\and
\IEEEauthorblockN{2\textsuperscript{nd} Moyi Tian}
\IEEEauthorblockA{\textit{Dept. of Applied Mathematics} \\
\textit{University of Colorado Boulder}\\
Boulder, CO, USA\\
ORCID: 0000-0001-8166-7410}
\and
\IEEEauthorblockN{3\textsuperscript{rd} Nancy Rodr\'iguez}
\IEEEauthorblockA{\textit{Dept. of Applied Mathematics} \\
\textit{University of Colorado Boulder}\\
Boulder, CO, USA\\
ORCID: 0000-0003-1023-9659}
}

\maketitle

\begin{abstract}
Social media increasingly shapes everyday life, serving as both a central venue for discussion of major events and a space where online collective behavior can spill over into real-world activity, while artificial intelligence (AI) is likewise becoming increasingly influential across society. Understanding how different online communities respond to AI-related events, and disentangling the mechanisms underlying the development of such engagement, are therefore increasingly important for studying information spreading and the societal reception of AI. Our work addresses the limited connection between empirical studies of AI-related online engagement and mathematical modeling of group-level spreading dynamics. We develop a framework to collect and organize empirical Bluesky activity into distinct user groups, then use a network-based dynamics model to investigate the mechanisms underlying their engagement. We use the release of DeepSeek R1 as a case study under two complementary grouping schemes: AI-related communities and academic disciplines. For each group, we fit a network-based Susceptible--Infected--Recovered (SIR)-type model augmented with an exogenous engagement term, allowing us to quantify the relative strengths of endogenous network-driven spreading and direct external response. Across groups, we find that strong direct responses to the event do not necessarily coincide with strong network-driven propagation, revealing distinct engagement patterns that may be obscured by aggregate activity alone.
\end{abstract}


\section{Introduction}
\label{sec:intro}

Social media platforms have become major channels through which large populations encounter, discuss, and react to major public events in near real time, providing a rich and increasingly accessible record of collective human behavior at population scale. According to \cite{DataReportal}, the identities of active social media users reached $69.9 \%$ of the global population in April 2026, representing an increase of $5.4 \%$ over the preceding year. Within this landscape, artificial intelligence (AI) has become an increasingly prominent topic of public attention. Large language models (LLMs) have been projected to affect a substantial share of tasks in the workforce \cite{eloundou2024gpts}, while public awareness and use of AI have increased sharply in recent years, for example, in recent U.S. surveys \cite{pew2026aiviews}. Understanding how online populations engage with AI-related developments is therefore of growing interest both for the study of information diffusion and for understanding the societal reception of AI.

Previous work has approached social media dynamics from complementary empirical, network, and mathematical perspectives. Empirical studies have characterized platform use \cite{Joinson_2008, Smock_2011} and shown that influence along online ties produces real-world behavioral change \cite{Bond_2012aa}, while network studies have shown that features such as tie strength \cite{Bakshy_2012}, neighborhood structure \cite{Ugander_2012}, community structure \cite{cinelli_2021}, and cascade topology \cite{Goel_2016} can substantially shape information diffusion. Mathematical approaches have also modeled online activity as a spreading process, including epidemic-inspired models of collective bursts and event-driven diffusion \cite{Crane_2008, Takayasu_2015}, as well as point-process and growth-based approaches that characterize the timing and volume of online engagement \cite{alipour_2024}. Extensions have distinguished endogenous network-driven spreading from exogenous influence \cite{Myers_2012}, used multivariate point processes to quantify directed influence between distinct online communities \cite{Zannettou_2018}, and connected compartmental dynamics with self-exciting point processes \cite{rizoiu2018sirhawkes}. Although epidemic-inspired models provide a useful abstraction for social spreading, social contagion should not be assumed to follow the same mechanisms as biological disease transmission and may involve broader behavioral and contextual effects \cite{Porter_2016}. We therefore use epidemic-inspired compartmental modeling as a parsimonious framework for representing engagement dynamics rather than as a literal model of disease transmission.

Recent studies have examined public discourse surrounding generative AI through sentiment, topic, and network analyses \cite{Qi_2024, Patel_2026}. Related work has demonstrated substantial differences across user groups, including occupation-dependent attitudes and discussion patterns toward generative AI \cite{Miyazaki_2024aa}, while broader studies of social-media events have shown that different communities may respond differently to the same external event in both activity and network structure \cite{Lu_2014aa}. Community-specific epidemic models have also been used to characterize differences in spreading dynamics across follower networks \cite{cava_2023}. These findings highlight that online populations are not homogeneous and that aggregate platform-level behavior may obscure important group-level heterogeneity. However, little work has examined how distinct communities respond dynamically to the same AI-related event while jointly considering temporal engagement, social-network structure, and endogenous versus exogenous mechanisms.

Our work is an initial effort to address this gap, studying group-level engagement with AI-related topics on Bluesky, a platform with a rapidly growing user base that has also attracted increasing attention as an accessible source of social media data \cite{Smith2026, Quelle_2026}. We collect longitudinal posting activity for users in selected Bluesky communities and align their activity with independently dated releases of major AI model families, using these releases as external events around which engagement dynamics can be examined. Among these events, we focus on the release of DeepSeek R1 as a case study because it is associated with a pronounced, near-simultaneous burst of on-topic discussion across a broad range of groups. We examine this common external event under two complementary grouping schemes: one based on AI-related communities and the other on academic disciplines. Each scheme defines multiple groups by a distinct criterion, allowing us to examine whether engagement dynamics differ across groups and across ways of organizing the online population. For each group, we construct a group-specific directed follow network and use members' temporal on-topic posting activity to reconstruct their Susceptible--Infected--Recovered (SIR) states over time, where the ``Infected'' state denotes active engagement rather than literal infection. We then fit these data using a network-based SIR-type model augmented with an exogenous engagement term, which accounts for activity stimulated directly by the external event rather than through the observed follow network. Recovery and exogenous engagement rates are estimated from data, leaving the network-driven spreading rate as the model's primary fitted parameter. This framework allows us to compare groups not only by the magnitude of their overall response, but also by the relative strengths of two mechanisms underlying that response: direct reaction to the external event and network-driven spreading within the group. We summarize these mechanisms using an effective network reproduction number, interpreted as endogenous spreading potential, and the ratio of event-driven to ambient exogenous engagement, which quantifies the strength of the direct external response. We find that these two measures are not necessarily aligned across groups: groups with strong externally driven responses do not always exhibit correspondingly strong endogenous spreading potential.

The rest of the paper is organized as follows: Section~\ref{sec:data_collection} describes our data pipeline, Section~\ref{sec:framework} characterizes the empirical data and presents the model and estimation procedure, Section~\ref{sec:result} reports results for the DeepSeek R1 case study, and Section~\ref{sec:conclusion} discusses implications and limitations.

\section{Data Collection and Processing}
\label{sec:data_collection}

\begin{figure}[!htbp]
    \centerline{\includegraphics[width=.95\linewidth]{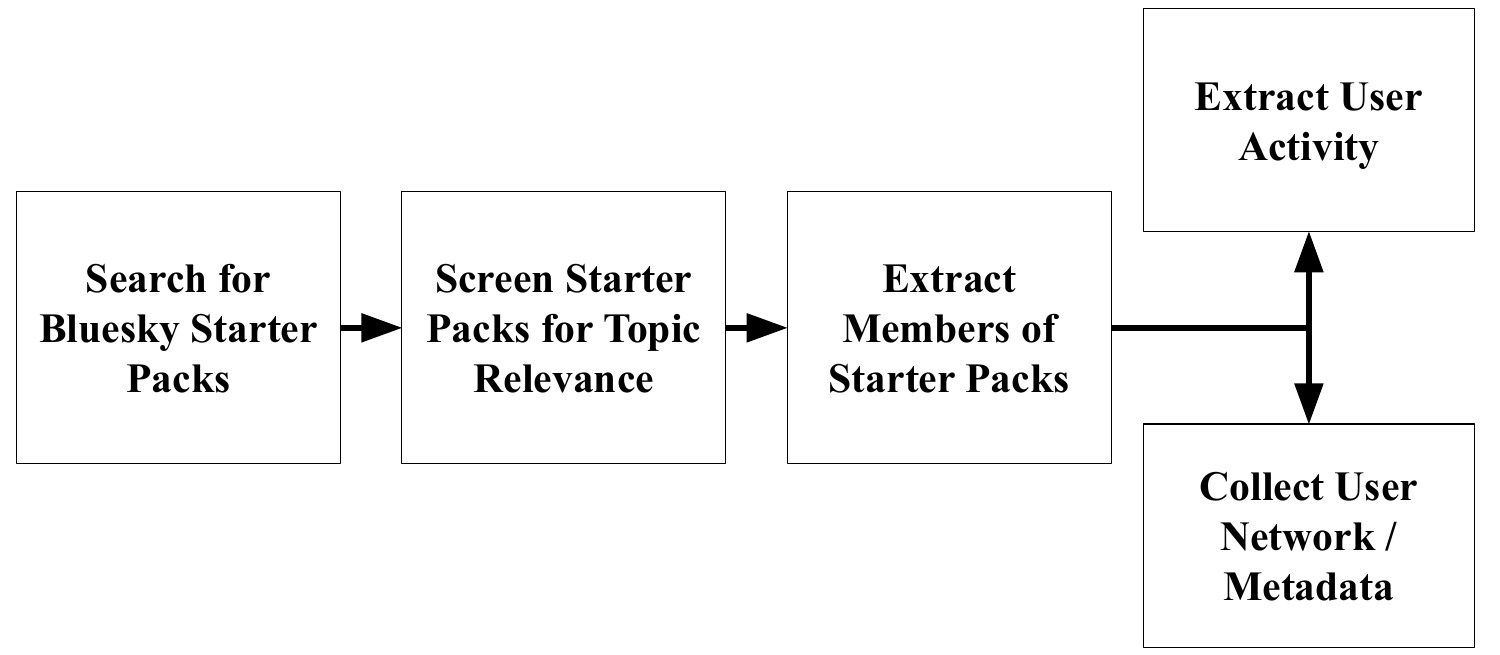}}
    \caption{Data acquisition and processing pipeline.}
    \label{fig:pipeline}
\end{figure}

To acquire our data from Bluesky, we developed Python scripts to collect publicly available information through the official Bluesky API and the Personal Data Servers (PDSs) associated with users' Decentralized Identifiers (DIDs) between January 2025 and August 2026 \cite{atproto2026api}. We use Bluesky starter packs as a discovery mechanism for identifying topically related groups. Previous work has shown that starter-pack membership exhibits substantial community structure rather than representing arbitrary account collections \cite{Smith2026}. Building on this observation, we use topic-specific starter packs to identify user groups whose network ties and posting activity are subsequently tracked longitudinally. We first query Bluesky for posts containing the phrase ``starter pack'' together with keywords related to the topic group of interest, e.g.\ ``AI'', ``machine learning'', or ``LLM''. These posts are then screened for valid embedded links to Bluesky starter packs. Figure~\ref{fig:pipeline} summarizes the overall data acquisition and processing workflow: starter-pack discovery and screening, network construction, and activity collection (Sections~\ref{subsec:starter_pack}--\ref{subsec:activity_data}).

\subsection{Starter Pack Classification}
\label{subsec:starter_pack}

We organized users under two complementary grouping schemes, AI-related communities and academic disciplines, each collected through its own discovery track. Each of the eleven academic-discipline datasets (e.g.\ Physics, Marine Biology, Political Science) was discovered using a keyword set curated for that domain, e.g.\ ``physics, astrophysics, physicist'' for the Physics dataset. The AI-topic datasets, by contrast, were all discovered using one shared AI keyword set, containing AI-related words such as ``AI'', ``LLM'', and ``machine learning'', and were sorted into six groups. We then passed the name and description of each candidate pack found to a zero-shot text classifier, \texttt{deberta-v3-large-zeroshot-v2.0}, run in multi-label mode against a set of candidate labels describing the target topic (e.g., ``Marine biology or oceanography community'' paired with ``Not related to marine biology or oceanography'' for the Marine Biology domain). A pack is retained only if the classifier's highest-scoring label is positive and exceeds a confidence threshold of 0.6, or if confirmed on-topic by human review; every other pack is discarded. Because packs with short descriptions produce unreliable classifier confidence, for those packs we always manually review rather than relying on the classifier alone.

For the AI-related packs, discovery was designed to be broad, retaining anything AI-adjacent, so the pool spans several distinct communities, e.g., AI research and engineering, AI ethics and policy, generative-AI enthusiasts, and AI critics. We reviewed this pool manually, since some packs rely on world knowledge the classifier lacks, and some creative communities oppose AI without stating it explicitly. We then sorted the retained packs by name and description into six groups: AI Research, for packs about the technical and professional practice of AI/ML; Non-AI Artists, for creative communities with an explicit no-AI stance; AI Artists, for creative communities that use generative AI; AI Activists (anti), for packs opposing AI; and AI Activists (pro), for packs advocating wider AI use or discussion. Remaining packs went to AI General Topic, which covers AI-adjacent packs with no particular stance.

The resulting groupings are mostly but not entirely disjoint, since a pack can fit more than one grouping, and packs in different groupings can share members. The overlap is modest: 8.9\% of AI Activists (pro) and 3.9\% of AI Activists (anti) users also appear in AI Research, and 14.3\% and 9.0\% respectively in AI General Topic, leaving the large majority of each group (83.7--90.0\%) unique to it.

\subsection{Network Construction}
\label{subsec:net_construct}

For each retained pack, we query the Bluesky API for its member DID list. For each member, we then collect their handle, display name, follower count, follow count, and their full list of accounts followed and blocked, up to a cap of 100,000 per user. If a user's list exceeds this cap, we flag the record as truncated and query that user directly against the network to resolve their in-network follows. This step produces a directed follow graph restricted to accounts belonging to at least one retained pack. Truncation is rare in practice: across all crawled datasets, only 13 of 43,582 user records (0.03\%) were flagged, concentrated in the largest datasets by user count.

\subsection{Activity Data Collection}
\label{subsec:activity_data}

For each user in the network, we queried their post history for the fixed window between January 1, 2025 and August 1, 2026. This information includes each post's raw text and its timestamp. To identify which of these posts constitute on-topic engagement with a given event, we apply a keyword filter to the full text of each post: a post is labeled on-topic if it contains at least one keyword from an event-specific keyword list, matched case-insensitively against the post's raw text. This activity-level keyword filter is applied independently of the discovery keywords mentioned at the beginning of this section, which are used only to find and classify starter packs during network construction.

The code used to implement the data pipeline, as well as the modeling experiments and analyses presented in this work, is available at https://github.com/AurumExile/Bluesky-Engagement-Research-Public.

\section{Data Analysis and Modeling Framework}
\label{sec:framework}

Section~\ref{subsec:prelim_data_analysis} characterizes the group networks and the temporal dynamics of AI-related engagement that motivate our model, and Section~\ref{subsec:math_model} defines the model and its estimation procedure.

\subsection{Empirical Characterization}
\label{subsec:prelim_data_analysis}

We characterize the datasets and the empirical properties of engagement that motivate the model's design. Table~\ref{tab:network_characteristics} summarizes the directed follow graph $A$ crawled for each of the seventeen communities, split into two grouping schemes as described in Section~\ref{sec:data_collection}. For each group, $N$ is the number of nodes retained in $A$; $\langle k \rangle$ is the mean degree (number of edges per node); $k_{\text{in}}^{\max}$ and $k_{\text{out}}^{\max}$ are the largest in-degree (most-followed account) and out-degree (account following the most others) observed in the graph; $C$ is the average local clustering coefficient, computed on the undirected projection of $A$ with each directed edge symmetrized before applying the standard triadic definition. The value $\lambda_{\max}(A)$ is the spectral radius of the adjacency matrix, reported here because it reappears later in Section~\ref{subsec:math_model} as the structural amplification term in the endogenous spreading potential. Network sizes range from several hundred to several thousand nodes, with mean degree $\langle k \rangle$ between $34$ and $211$, indicating that these are internally well-connected, rather than incidentally connected, populations.

\begin{table}[!htb]
\centering
\footnotesize
\setlength{\tabcolsep}{2pt}
\caption{Network characteristics of each group's directed follow graph $A$}
\label{tab:network_characteristics}
\setlength{\tabcolsep}{1.5pt}
\begin{tabular}{lrrrrr@{\hspace{8pt}}r}
\toprule
Group & $N$ & $\langle k \rangle$ & $k_{\text{in}}^{\max}$ & $k_{\text{out}}^{\max}$ & \multicolumn{1}{r@{\hspace{3pt}}}{$\lambda_{\max}(A)$} & $C$ \\
\midrule

\multicolumn{7}{l}{\textit{AI-related communities}} \\
\addlinespace[1pt]
Non-AI Artists          & 2{,}398 &  69.93 &   544 & 1{,}353 & 103.39 & 0.454 \\
AI Research             & 1{,}826 & 103.33 &   753 &   972 & 155.33 & 0.525 \\
AI General Topic        & 1{,}322 &  99.26 &   539 & 1{,}049 & 132.13 & 0.553 \\
AI Activists (anti)     &    588 &  35.22 &   211 &   261 &  37.88 & 0.559 \\
AI Activists (pro)      &    349 &  34.34 &   168 &   175 &  49.29 & 0.537 \\
AI Artists              &    277 &  39.47 &   125 &   179 &  49.71 & 0.751 \\

\midrule
\multicolumn{7}{l}{\textit{Academic disciplines}} \\
\addlinespace[1pt]
Public Health/Epidemiology & 3{,}535 & 178.45 & 1{,}443 & 2{,}450 & 361.92 & 0.507 \\
Economics               & 3{,}302 & 211.29 & 1{,}752 & 1{,}897 & 345.20 & 0.483 \\
Political Science       & 3{,}060 & 163.78 & 1{,}042 & 2{,}106 & 287.97 & 0.451 \\
Clinical/General Medicine & 2{,}973 & 133.31 &   1{,}184 & 2{,}611 & 249.41 & 0.547 \\
Computer Science        & 2{,}822 &  60.89 & 1{,}436 &   621 & 108.84 & 0.520 \\
Sociology/Anthropology  & 2{,}486 & 127.89 &   809 & 1{,}442 & 230.07 & 0.502 \\
Genetics/Biotech        & 2{,}460 & 147.08 & 1{,}041 & 1{,}891 & 239.05 & 0.510 \\
Engineering             & 2{,}411 &  47.12 &   579 &   478 &  80.20 & 0.548 \\
Physics                 & 1{,}806 &  76.93 &   919 &   700 & 145.65 & 0.521 \\
Plant Science           & 1{,}181 & 128.18 &   673 & 1{,}109 & 191.87 & 0.584 \\
Marine Biology          & 1{,}049 & 116.30 &   548 &   789 & 158.63 & 0.551 \\
\bottomrule
\end{tabular}
\end{table}

Across both schemes, group size $N$ spans roughly an order of magnitude, from 277 (AI Artists) to 3{,}535 (Public Health/Epidemiology), while mean degree $\langle k \rangle$ varies nearly sixfold without scaling proportionally to $N$: several of the smaller AI-related communities (e.g.\ AI Research, $N=1{,}826$, $\langle k \rangle = 103.33$) are comparably or more densely connected than larger academic groups. The maximum in- and out-degrees also diverge substantially within groups (Non-AI Artists: out-degree 1{,}353 vs. in-degree 544; Computer Science: the reverse, in-degree 1{,}436 vs. out-degree 621), indicating that hubs are heavy followers in some communities and heavily followed accounts in others. $\lambda_{\max}(A)$ tracks $N$ and $\langle k \rangle$ loosely, underscoring that this eigenvalue captures structural amplification not fully reducible to either raw size or average connectivity alone. Clustering coefficients $C$ sit within a comparatively narrow band (0.451--0.751) despite the wide variation in every other column, indicating that local triadic structure is a shared property across all seventeen communities.

\begin{figure*}[!htbp]
    \centerline{\includegraphics[width=.85\linewidth]{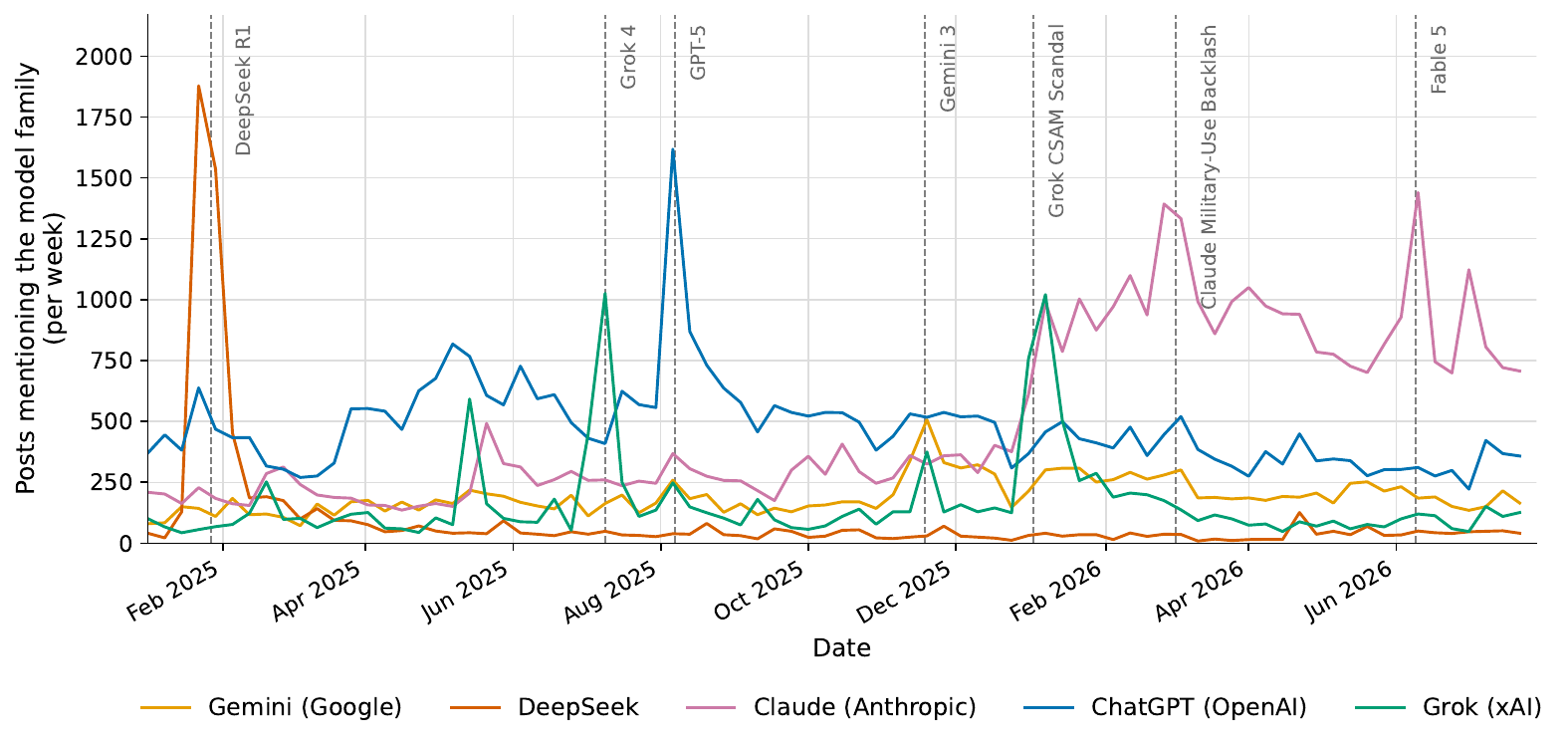}}
    \caption{Weekly post volume mentioning five major AI model families across all crawled datasets (Jan.\ 2025--Aug.\ 2026). Dashed lines mark independently dated model releases (DeepSeek R1, Grok 4, GPT-5, Gemini 3, Fable 5) and controversies (Grok CSAM scandal, backlash over Claude's military use). Most events trigger bursts that decay to baseline; Claude instead shows a persistently elevated baseline from early 2026, with later events adding spikes.}
    \label{fig:time_series_general}
\end{figure*}

Beyond network structure, engagement with AI-related topics is punctuated by sharp, short-lived bursts rather than smooth organic growth. Figure~\ref{fig:time_series_general} tracks posts mentioning five major AI model families, identified by per-family keyword filters (e.g.\ ``deepseek'' for DeepSeek; ``chatgpt, gpt-5, gpt5'' for ChatGPT), aggregated across all crawled datasets from January 2025 to August 2026. We compare the weekly post volume against seven independently documented events: model releases (DeepSeek R1, Grok 4, GPT-5, Gemini 3, Fable 5) and non-release controversies (the Grok CSAM scandal, the backlash over Claude's military use).

Four of the five families show event-driven spike-and-decay dynamics. DeepSeek produces the tallest peak in the series at the DeepSeek R1 launch, then decays to a low, flat baseline with no comparable resurgence. Grok spikes sharply at both the Grok 4 release and the Grok CSAM scandal, returning toward baseline after each. Gemini remains comparatively muted, with only a modest, short-lived rise at Gemini 3. ChatGPT maintains the highest baseline until early 2026 and shows its largest spike at the GPT-5 launch, returning to that baseline afterward. Claude departs from this pattern in a way directly relevant to our modeling choices: its volume rises in early 2026, before either Claude-specific event, and stabilizes at a persistently elevated baseline, on which the military-use backlash and the Fable 5 release appear as additional spikes.

More broadly, this recurring structure, an identifiable exogenous trigger followed by a burst of engagement and subsequent decay, motivates representing engagement as an event-driven process with a discrete exogenous shock term rather than as smooth continuous growth (Section~\ref{subsec:math_model}). It also justifies treating individual events, such as the DeepSeek R1 case study in Section~\ref{sec:result}, as tractable, self-contained units of analysis.

\subsection{Mathematical Model}
\label{subsec:math_model}

We model engagement with a topic as a network-based, discrete-time SIR-type process with exogenous forcing, running over each group's directed follow graph. We retain the conventional SIR terminology for consistency with the classical compartmental framework, while interpreting infection and recovery in this application as entry into and exit from active topic engagement. This framing is also supported by Rizoiu et al. \cite{rizoiu2018sirhawkes}, who show that the expected rate of new infections in a stochastic SIR process equals the event rate of an extended Hawkes process after marginalizing over recovery events. This connection provides additional motivation for representing discrete, timestamped engagement activity on social media with a compartmental framework. We use an SIR rather than a Susceptible--Infected--Recovered--Susceptible (SIRS) structure so that the Recovered state is absorbing in the model, focusing on the spread and scale of a single event-driven response within communities rather than repeated long-term engagement on a topic.

We now formalize this construction at the level of individual users and their posting activity. A node $i$ is assigned to the Infected state ($I$), representing active engagement, at time bin $t$ if it posted on-topic content, as identified by an event-specific keyword filter, in any of the $\tau$ most recent bins, including bin $t$. A new on-topic post resets this window and returns the node to $I$ if it had already reached $R$. Once $\tau$ bins have elapsed without further on-topic posts, $i$ transitions to Recovered ($R$). Time is divided into bins of width $\Delta t$; results in this paper use $\Delta t = 1$ hour and $\tau = 24$ bins, corresponding to a 1-day active-engagement window.

Let $A$ be the $n \times n$ directed adjacency matrix of the follow graph, where $n$ denotes the number of users in the group and $a_{ij} = 1$ if $i$ follows $j$ and $0$ otherwise, so that on-topic content posted by $j$ is visible to $i$. To estimate model parameters from the discrete-time trajectories reconstructed above, we fit an individual-based mean-field approximation to the process. Each node's discrete state is approximated by continuous variables $S_i(t), I_i(t), R_i(t) \in [0,1]$, interpreted as the probabilities that node $i$ is in each compartment at time $t$, with $S_i(t) + I_i(t) + R_i(t) = 1$. These evolve in continuous time as
\begin{equation}
\left\{
\begin{aligned}
    \frac{dS_i}{dt} &= -\beta_w \, S_i \, e_i(t) - \mu(t)\, S_i, \\
    \frac{dI_i}{dt} &= \beta_w \, S_i \, e_i(t) + \mu(t)\, S_i - \gamma \, I_i, \\
    \frac{dR_i}{dt} &= \gamma \, I_i,
\end{aligned}
\right.
\label{eq:math_model}
\end{equation}
where $e_i(t) = \sum_j a_{ij} I_j(t)$ is the expected number of accounts $i$ follows that are currently actively engaged, $\beta_w$ is the per-edge network-driven spreading rate within sub-window $w$, $\gamma$ is the rate at which nodes exit active engagement, and $\mu(t)$ is the time-dependent exogenous engagement rate acting independently of network exposure $e_i(t)$. During estimation, $\gamma$ and the function $\mu(t)$ are fixed from data, whereas $\beta_w$ is piecewise constant fitted independently within each adaptive sub-window, allowing network-driven spreading strength to vary over the course of the event.

Because a new on-topic post extends a node's time in $I$, or returns it to $I$ after a lapse, the reconstruction above departs from the standard SIR-type model, in which the recovery rate is constant, the resulting time in $I$ is memoryless, and $R$ is absorbing: a node's total time in $I$ depends on its full pattern of re-engagement and can extend arbitrarily beyond a single $\tau$-window. These choices are made so that the observed $I(t)$ reflects all active engagement. The continuous-time compartmental model in Eq.~\eqref{eq:math_model}, fit with a single, constant recovery rate $\gamma$, is therefore best understood as an effective mean-field approximation of this aggregate process, where $\gamma$ represents the population-averaged rate at which nodes exit active engagement, estimated empirically as described below.

For a state $X \in \{S, I, R\}$, let $n_X$ be the total number of node-bin observations in state $X$, pooled over all nodes and all bins in the relevant window, and let $n_{X \to Y}$ count those in which the node is in state $Y$ in the next bin ($Y = \lnot X$ denotes any state other than $X$). The ratio $n_{X \to Y}/n_X$ is then the empirical per-bin probability of transitioning from $X$ to $Y$. Rather than fitting $\gamma$ to the aggregate engagement curve, we set $\gamma = n_{I \to \lnot I} / n_I$, the observed per-bin exit rate from the $I$ compartment, computed over the event window for fitting and, analogously, over the calibration window when estimating $\mu_{\mathrm{ambient}}$ (below). Episodes still open at the window's end contribute to $n_I$ without an exit, so they are retained rather than dropped. This estimate is informative only when nodes post again before their $\tau$-window expires: if every engaged node posts only once, then each spends exactly $\tau$ bins in $I$, and $\gamma = 1/\tau$ reflects our choice of $\tau$ rather than observed behavior. We therefore report a group/event fit as reliable only if at least $m$ users post more than once in the event window. We call this threshold, $m$, the multi-post-user reliability gate. Fits below this gate are still shown but flagged. The same threshold $m$ also sets the minimum-data requirement for estimating $\mu_{\mathrm{event}}$ and the stopping criterion for the sub-window growth (below). We use $m = 15$, low enough to retain most communities as reliable but high enough that $\gamma$, $\mu$, and sub-window widths reflect sufficient data.

We next specify the exogenous forcing $\mu(t)$. \label{sec:mu} Rather than fitting it jointly with $\beta_w$, we estimate it directly from data and hold it fixed during fitting. Let $t_e$ denote the beginning of the fixed 24-hour window corresponding to the externally dated event (e.g., a model release or the date of viral news). We set
\begin{equation*}
    \mu(t)=
        \begin{cases}
        \mu_{\mathrm{event}},
        & t_e \leq t < t_e+24\ \mathrm{hours},\\
        \mu_{\mathrm{ambient}},
        & \text{otherwise},
        \end{cases}
\end{equation*}
where $\mu_{\mathrm{ambient}}$ is the baseline exogenous engagement rate and $\mu_{\mathrm{event}}$ is the exogenous engagement rate during the event window. This exogenous/endogenous split echoes the background-rate term in point-process models of online diffusion, where event intensity is a background rate plus a self-exciting component driven by prior events \cite{rizoiu2018sirhawkes}. Unlike that setting, where the background rate is typically zero because a cascade is assumed to start from a single triggering event, we retain a nonzero $\mu$, since topic engagement in our communities is plausibly seeded both by network exposure and by sources outside the follow graph, such as mainstream coverage of the model's release. We estimate $\mu_{\mathrm{ambient}}$ once per group/event pair from a period free of notable events outside the event window (pre-event where possible, otherwise post-event), which we call the calibration window. Since network exposure $e_i(t)$ is observed, we fit the discrete-time form of Eq.~\eqref{eq:math_model} over the calibration window by least squares, with a single constant spreading rate and a constant exogenous rate. We keep only the latter as $\mu_{\mathrm{ambient}}$, so network-driven conversions are absorbed by the spreading term rather than counted as exogenous. Groups with no on-topic activity in this window have $\mu_{\mathrm{ambient}} = 0$, leaving the event-to-ambient ratio undefined, and we report only their raw engagement statistics. To isolate the event's direct effect from network spreading, we estimate $\mu_{\mathrm{event}}$ within the event window as the $S$-to-$I$ transition probability only among susceptible nodes that follow no currently engaged account ($e_i (t) = 0$). If fewer than $m$ such node-bin observations are available, $\mu_{\mathrm{event}}$ is set to $\mu_{\mathrm{ambient}}$.

With $\gamma$ and $\mu$ fixed as above, the only remaining parameter is $\beta_w$, which we fit independently for each contiguous, non-overlapping sub-window $w$ of the event period, since network conditions can shift over the course of a sustained event. Sub-windows are not tied to a calendar unit. Each sub-window grows in one-day increments until it contains at least $m$ on-topic posts, so its width adapts to local activity rather than imposing a fixed resolution on both busy and quiet activity periods. Within each sub-window, we numerically integrate the system from the observed state $(S_i, I_i, R_i)$ at its start using SciPy's \texttt{solve\_ivp} with the implicit Backward Differentiation Formula method~\cite{virtanen2020scipy}, and choose $\beta_w$ by bounded scalar minimization of the squared error against the observed $I(t)$ over $\beta_w \in [0,\, 10\gamma/\lambda_{\max}(A)]$. The lower bound reflects that network-driven spreading rates cannot be negative; the upper bound is discussed below.

For each sub-window $w$, we define its endogenous spreading potential and use it as the summary measure: $$R_0(w) = \frac{\beta_w}{\gamma}\,\lambda_{\max}(A),$$ where $\lambda_{\max}(A)$ is the spectral radius of the adjacency matrix $A$, which generalizes the well-mixed SIR threshold $\beta/\gamma$ to networked contact structure under this mean-field approximation \cite{PastorSatorras_2015, Mei_2017}. Without exogenous forcing, $R_0(w)=1$ separates subcritical spreading, in which engagement introduced through the network dies out, from supercritical spreading, in which it can initially amplify. Since the full model includes exogenous engagement, $R_0(w)<1$ does not imply that observed activity must vanish, as external stimulation can sustain engagement independently of the network-driven spreading. The upper search bound on $\beta_w$ corresponds to $R_0 (w) = 10$, well above any converged peak value in this study, so the interval does not constrain the fit. A fit pinned at this ceiling indicates a boundary solution rather than an interior optimum, and we do not report it as converged; among our groups, this occurred only for Physics. Because $\mu$ is fixed rather than jointly estimated, $\beta_w$ has a network-only interpretation, and $R_0(w)$ is the quantity we compare across communities and events.

Finally, we assess fit quality using the coefficient of determination between the observed aggregate engagement curve and the model's simulated trajectory,
$$R^2 = 1 - \frac{\sum_t \left(I(t) - \hat{I}(t)\right)^2}{\sum_t \left(I(t) - \bar{I}\right)^2},$$
where the sums run over all time bins $t$ in the event window, $I(t)$ is the observed fraction of nodes in the active-engagement state $I$ in bin $t$, $\hat{I}(t)$ is the fitted fraction at the same bins, and $\bar{I}$ is the mean of $I(t)$ over the full event window. An $R^2$ of 1 indicates a perfect match between simulated and observed engagement. Because $\hat{I}(t)$ comes from a nonlinear dynamical system rather than a linear projection, a poor fit can explain less variance than the constant prediction $\bar{I}$, so $R^2$ can be negative. We treat $R^2 < 0$ as evidence that the model captures no meaningful structure for that group/event pair. Because each sub-window is reseeded from the observed state, $R^2$ measures how closely the model tracks engagement between reseeding points rather than over the full event from a single initial condition. We therefore use it to screen fit quality, not to compare mechanisms across groups.

\section{Results}
\label{sec:result}

Our case study is the release of DeepSeek R1, an open-weight large language model released by the Chinese AI lab DeepSeek on January 20, 2025. The company reported that it performs competitively with leading closed frontier models at a small fraction of their disclosed training cost \cite{deepseekr1_2025}. This technical release drew immediate attention within machine-learning circles, but the broader public reaction it eventually provoked did not peak until a full week later: by January 27, 2025, DeepSeek's mobile app had overtaken ChatGPT atop Apple's U.S. App Store \cite{npr2025deepseek}, and the news was accompanied by a historic single-day sell-off in AI-linked technology stocks. Nvidia alone lost roughly \$590 billion in market value \cite{cnbc2025nvidia}, drawing coverage across mainstream and financial press \cite{washingtonpost2025deepseek}. Because Bluesky is a general-audience platform rather than one scoped to machine-learning practitioners, we anchor the $\mu_{\mathrm{event}}$ window described in Section~\ref{sec:mu} to January 27, the date of peak mainstream public attention, rather than the January 20 technical release date. This date is set externally, by app-store rankings, market reaction, and press coverage, rather than by our data. The launch is thus a useful common event for our analysis: it produced a pronounced, near-simultaneous spike in on-topic discussion across both AI-related and academic communities in our dataset. We analyze this event within a fixed window spanning January 1 to March 1, 2025, a 59-day span that captures the weeks of relative quiet leading up to the spike as well as its full decay. We identify on-topic engagement with this event via a case-insensitive keyword match against the literal strings \texttt{"deepseek"} and \texttt{"deep seek"} in each post's raw text.

We fit the network-based SIR-type model described in Section~\ref{subsec:math_model} independently to each group's follow graph, using the DeepSeek R1 launch as the shared event window. Physics, Plant Science, and Marine Biology recorded no on-topic activity in the calibration window, so $\mu_{\mathrm{ambient}} = 0$ and the event-to-ambient ratio is undefined; for these groups, we report only their raw engagement statistics. AI Artists, Non-AI Artists, and Clinical/General Medicine produced enough on-topic activity for a fit but fell below the $m=15$ multi-post-user reliability gate, so their reported estimates rest on thin repeat-engagement data. Section~\ref{subsec:AI_group_results} presents results for AI-related communities, Section~\ref{subsec:discipline_group_results} for academic disciplines, and Section~\ref{subsec:cross_group_patterns} discusses patterns across both grouping schemes.

\subsection{AI-Related Communities}
\label{subsec:AI_group_results}

\begin{figure}[!htb]
    \centerline{\includegraphics[width=0.95\linewidth]{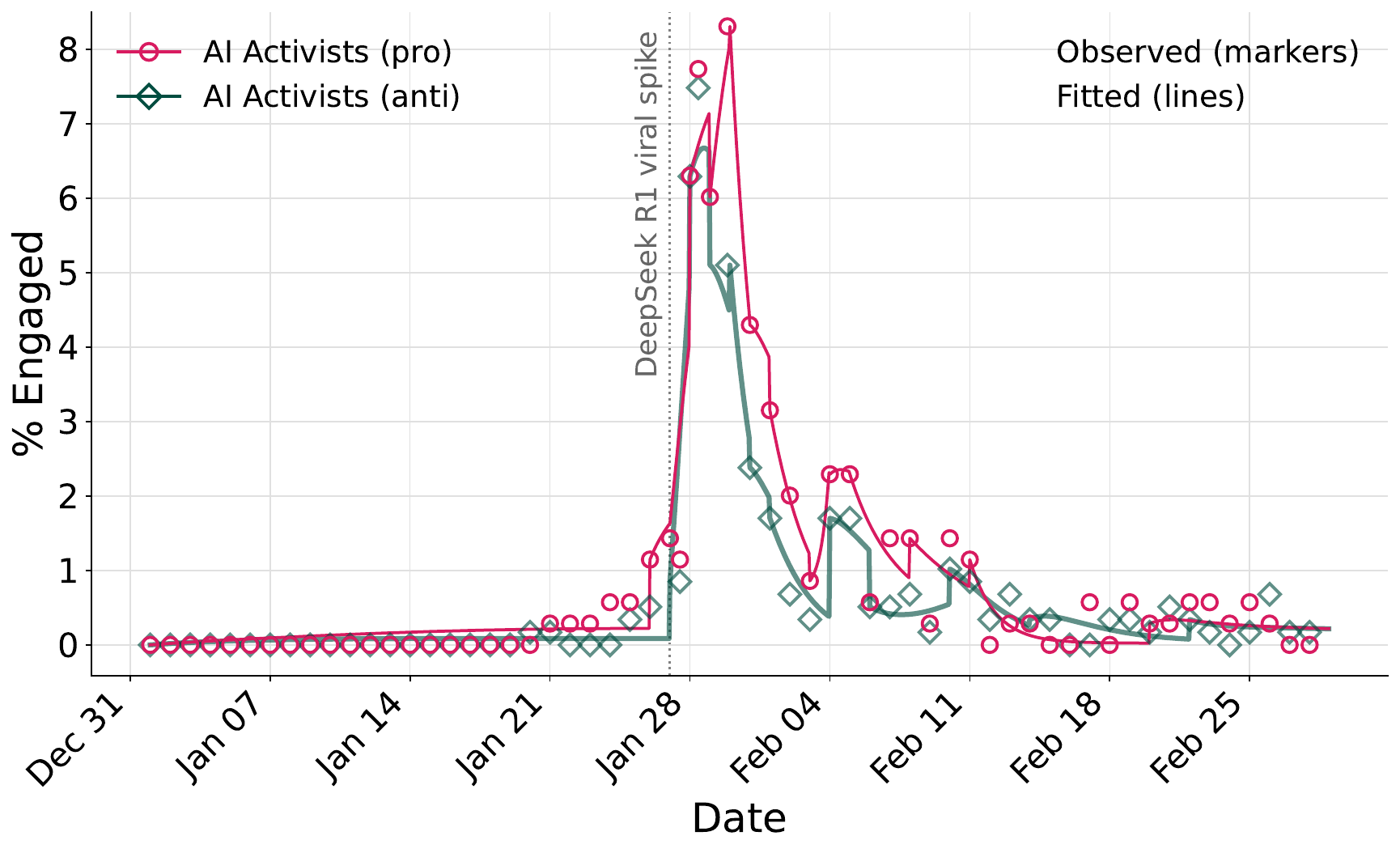}}
    \caption{Fitted vs.\ observed engagement ($I(t)$) for AI Activists (pro) and AI Activists (anti). The dotted gray line marks Jan.\ 27, 2025, the date of peak mainstream attention to DeepSeek R1. Visible discontinuities in the fitted trajectories arise because each adaptive sub-window is reseeded from the observed state at its boundary (Section~\ref{subsec:math_model}). Despite nearly identical exogenous ratios, the pro group reaches a taller, later peak and decays more slowly, which the model attributes to higher endogenous spreading potential (peak $R_0$ 1.806 vs.\ 1.084).}
    \label{fig:ai_topic_fit}
\end{figure}

Table~\ref{tab:deepseek-ai} summarizes the fitted model for all six AI-related communities, and Fig.~\ref{fig:ai_topic_fit} shows the two AI Activists groups, selected because of their contrasting trajectories despite nearly identical exogenous responses. Across the four reliably fit groups, $R^2$ ranges from $0.9477$ to $0.9859$, indicating consistently strong fits. The two groups below the reliability gate show markedly weaker fits ($0.5150$ for AI Artists, $0.7542$ for Non-AI Artists), consistent with our expectation that thinner repeat-engagement data yields less reliable estimates.

\begin{table*}[!htb]
\centering
\caption{Network-based SIR-type dynamics summary for the DeepSeek R1 launch among AI-related communities, sorted by peak $R_0$.}
\label{tab:deepseek-ai}
\small
\setlength{\tabcolsep}{4pt}
\begin{tabular}{lccccccc}
\toprule
Dataset & \% Engaged & Avg/User & Max/User & $\mu_{\mathrm{event}}/\mu_{\mathrm{ambient}}$ & $R^2$ & Lag (d) & Peak $R_0$ \\
\midrule
AI Activists (pro)       & 19.8\% & 5.06 & 50 & 111.7$\times$ & 0.9539 & +2.9 & 1.806 \\
AI General Topic                 & 21.1\% & 5.75 & 137 & 52.0$\times$ & 0.9859 & +1.5 & 1.427 \\
AI Research              & 10.7\% & 4.24 & 137 & 18.9$\times$ & 0.9731 & +1.6 & 1.367 \\
AI Activists (anti)      & 16.0\% & 3.57 & 50 & 109.6$\times$ & 0.9477 & +1.4 & 1.084 \\
\midrule
\textit{AI Artists}\textsuperscript{a}     & 4.3\% & 4.00 & 16 & 0.0$\times$ & 0.5150 & +1.5 & 1.152 \\
\textit{Non-AI Artists}\textsuperscript{a} & 1.2\% & 1.43 & 4 & 953.5$\times$ & 0.7542 & +1.6 & 0.386 \\
\bottomrule
\end{tabular}
\begin{flushleft}\footnotesize
Avg/User and Max/User: mean and maximum on-topic posts per engaged user over the full event window. Lag (d) and Peak $R_0$ both refer to the bin where observed $I(t)$ peaks: Lag is the number of days from the anchored event date (Jan.\ 27, 2025) to that bin, and Peak $R_0$ is $R_0(w)$ in the adaptive window containing it. \textsuperscript{a}Falls below the $m=15$ multi-post-user reliability gate (event window); shown for completeness only.
\end{flushleft}
\end{table*}

Figure~\ref{fig:ai_topic_fit} highlights a striking contrast between the two AI Activists groups. Despite reacting to the DeepSeek R1 spike with almost identical exogenous ratios ($\mu_{\mathrm{event}}/\mu_{\mathrm{ambient}} = 111.7\times$ for AI Activists (pro) versus $109.6\times$ for AI Activists (anti), a difference of under $2\%$), the two groups' subsequent trajectories diverge sharply. Both curves rise together through the two days following the spike, briefly converging near a shared local maximum, before AI Activists (pro) continues climbing to a taller, later global maximum, while AI Activists (anti) decays back toward baseline more quickly. This divergence is captured directly by the fitted model: peak $R_0$ differs by nearly $70\%$ between the two groups ($1.806$ for pro versus $1.084$ for anti), and pro's engagement peaks more than a day later than anti's ($+2.9$ vs.\ $+1.4$ days). Because the two groups' immediate exogenous reaction to the same event was effectively identical, the model attributes this divergence to a difference in endogenous spreading potential rather than in how strongly each group responded to the news itself, a tightly matched illustration of the dissociation between exogenous response and endogenous spreading.

Engagement intensity varies considerably within this single grouping scheme. Among the four reliable groups, the share of the network actively engaging ranges from $10.7\%$ (AI Research) to $21.1\%$ (AI General Topic), and average per-user posting volume ranges from $3.57$ to $5.75$ posts. Individual posting intensity and exogenous responsiveness do not necessarily move together: AI Research ties AI General Topic for the highest single-user post count observed in this set ($137$), yet records the lowest $\mu_{\mathrm{event}}/\mu_{\mathrm{ambient}}$ ratio of the four ($18.9\times$, versus $52$--$112\times$ for the remaining three), indicating that a community can contain highly prolific individual posters without a correspondingly strong collective exogenous reaction to the event. AI Artists, meanwhile, show no event-day conversions among users who followed no engaged account, so the measured $\mu_{\mathrm{event}}$ is zero (Section~\ref{sec:mu}), and all event-day engagement is attributed to network-driven spreading, consistent with its peak $R_0$ above $1$ ($1.152$). This group, along with Non-AI Artists, falls below our $m=15$ multi-post-user reliability gate, so estimates for both groups rest on very few engagements and should be read with caution as noted in Section~\ref{subsec:math_model}.

\subsection{Academic Disciplines}
\label{subsec:discipline_group_results}

\begin{figure}[!htb]
    \centerline{\includegraphics[width=0.95\linewidth]{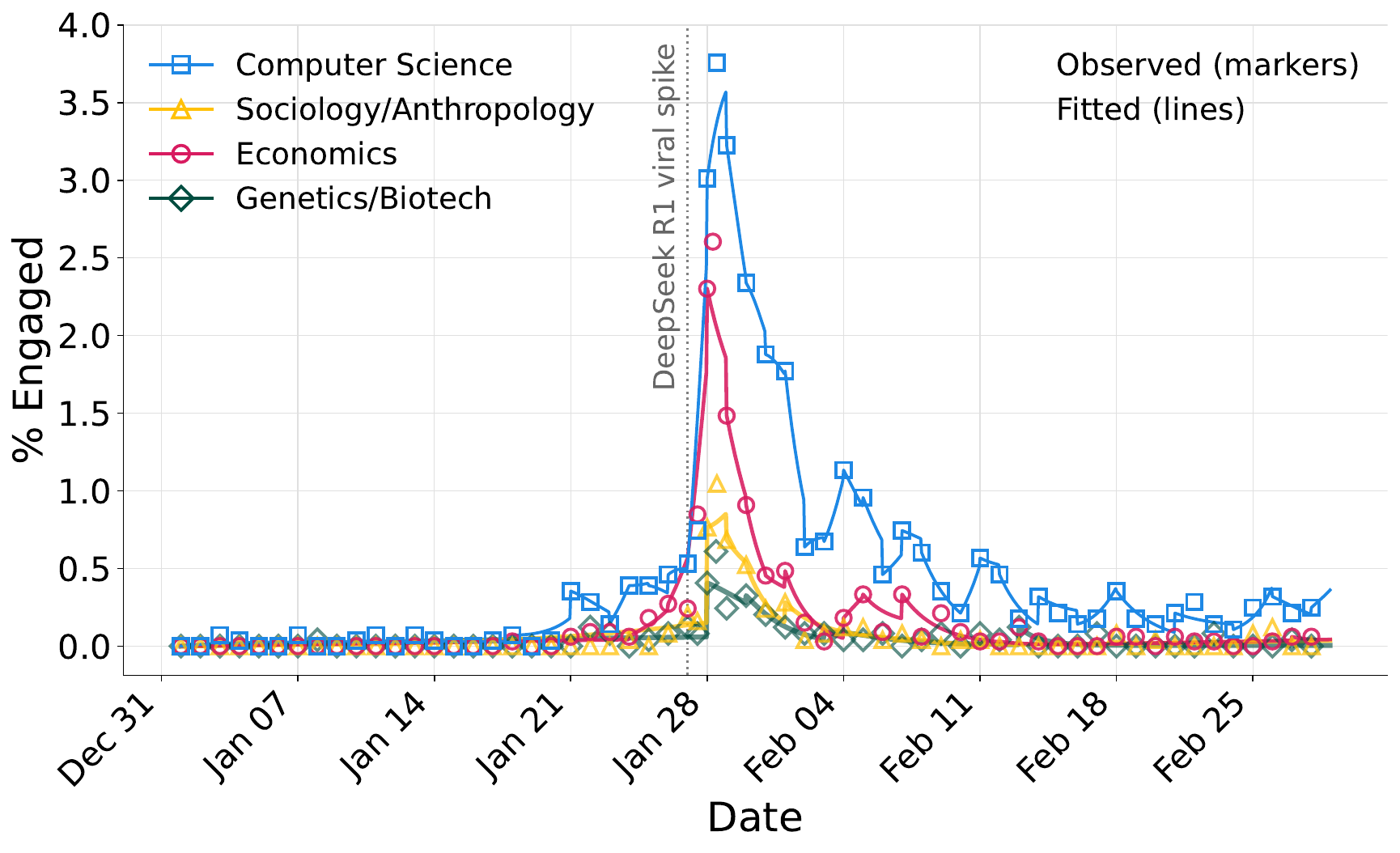}}
    \caption{Fitted vs.\ observed engagement ($I(t)$) for four academic disciplines; conventions follow Fig.~\ref{fig:ai_topic_fit}. Computer Science has the largest response and highest peak $R_0$. Economics has a $4.4\times$ larger exogenous ratio but a lower peak $R_0$. Sociology/Anthropology reaches a comparable peak $R_0$ despite an exogenous ratio $6.5\times$ larger, while Genetics/Biotech matches Computer Science's exogenous ratio yet spreads subcritically ($R_0<1$). The strength of a group's direct response to the event thus does not predict its network-driven spreading.}
    \label{fig:cs_fit}
\end{figure}

Table~\ref{tab:deepseek-domains} reports the same analysis for eleven academic disciplines, seven of which clear the reliability gate. Figure~\ref{fig:cs_fit} shows four that span the range of behaviors. Computer Science is the most active and well-fit discipline ($R^2=0.9827$), and its fitted curve also tracks a secondary rise in early February. Sociology/Anthropology has the largest exogenous ratio among reliably fit groups, yet a peak $R_0$ similar to Computer Science's. Economics has the second-largest exogenous ratio but a peak $R_0$ only slightly above the critical value of $1$. Genetics/Biotech is the only reliably fit group with peak $R_0 < 1$: its inferred network spreading is subcritical, so its engagement is sustained mainly by direct response to the event.

\begin{table*}[!htb]
\centering
\caption{Network-based SIR-type dynamics summary for the DeepSeek R1 launch among academic and professional communities, sorted by peak $R_0$.}
\label{tab:deepseek-domains}
\small
\setlength{\tabcolsep}{4pt}
\begin{tabular}{lccccccc}
\toprule
Dataset & \% Engaged & Avg/User & Max/User & $\mu_{\mathrm{event}}/\mu_{\mathrm{ambient}}$ & $R^2$ & Lag (d) & Peak $R_0$ \\
\midrule
Computer Science           & 9.7\% & 4.20 & 97 & 77.5$\times$ & 0.9827 & +1.5 & 1.456 \\
Sociology/Anthropology     & 2.7\% & 1.93 & 11 & 505.0$\times$ & 0.8531 & +1.5 & 1.402 \\
Political Science          & 4.0\% & 2.64 & 49 & 110.4$\times$ & 0.7520 & +1.3 & 1.276 \\
Engineering                & 7.3\% & 3.48 & 137 & 116.4$\times$ & 0.9702 & +1.5 & 1.270 \\
Public Health/Epidemiology & 3.2\% & 2.26 & 27 & 147.3$\times$ & 0.9441 & +1.5 & 1.214 \\
Economics                  & 5.1\% & 2.46 & 26 & 341.6$\times$ & 0.9516 & +1.3 & 1.034 \\
Genetics/Biotech           & 2.0\% & 1.54 & 5 & 76.1$\times$ & 0.7347 & +1.5 & 0.744 \\
\midrule
\textit{Clinical/General Medicine}\textsuperscript{a} & 1.6\% & 2.31 & 27 & 60.9$\times$ & 0.5905 & +1.5 & 1.103 \\
\textit{Physics}\textsuperscript{b}                 & 1.1\% & 1.47 & 3  & n/a & n/a & n/a & n/a \\
\textit{Plant Science}\textsuperscript{b}             & 0.6\% & 1.57 & 5  & n/a & n/a & n/a & n/a \\
\textit{Marine Biology}\textsuperscript{b}            & 0.4\% & 1.25 & 2  & n/a & n/a & n/a & n/a \\
\bottomrule
\end{tabular}
\begin{flushleft}\footnotesize
Column definitions and \textsuperscript{a} follow Table~\ref{tab:deepseek-ai}. \textsuperscript{b} No on-topic activity during the calibration window, so $\mu_{\mathrm{ambient}} = 0$ and the event-to-ambient ratio is undefined; model estimates are omitted and raw statistics shown for completeness.
\end{flushleft}
\end{table*}

Across the seven reliably fit disciplines, engagement rate ranges from $2.0\%$ (Genetics/Biotech) to $9.7\%$ (Computer Science), and the largest single-user post count reaches as high as $137$ in Engineering, matching the highest values observed among the AI-related communities and indicating that highly prolific individual posters are not unique to AI-related groups. Fit quality is generally strong but not uniform: five of the seven reliable disciplines record $R^2$ between $0.94$ and $0.98$, while Political Science ($0.7520$) and Genetics/Biotech ($0.7347$) fit visibly less well despite clearing the same gate, indicating that the reliability gate ensures sufficient data for estimation without guaranteeing strong fit.

Computer Science and Sociology/Anthropology reach comparable peak $R_0$ ($1.456$ vs.\ $1.402$) even though Sociology/Anthropology has the largest exogenous ratio among reliably fit groups, $6.5$ times that of Computer Science ($505.0\times$ vs.\ $77.5\times$); its engagement is nonetheless far smaller, with a lower engagement rate ($2.7\%$ vs.\ $9.7\%$) and fewer posts per engaged user ($1.93$ vs.\ $4.20$). Economics, with the second-largest exogenous ratio ($341.6\times$), reaches a peak $R_0$ only slightly above the critical value of $1$ ($1.034$), well below Sociology/Anthropology's despite a comparably strong exogenous response. Genetics/Biotech shares Computer Science's exogenous ratio almost exactly ($76.1\times$) yet is the only reliably fit group with subcritical spreading (peak $R_0=0.744$).

\subsection{Patterns Across Groups}
\label{subsec:cross_group_patterns}

Table~\ref{tab:deepseek-ai} and Table~\ref{tab:deepseek-domains} together allow a direct comparison of exogenous response and endogenous spreading potential across all eleven reliably fit groups in this study, spanning both grouping schemes. The $\mu_{\mathrm{event}}/\mu_{\mathrm{ambient}}$ ratio spans more than an order of magnitude, from $18.9\times$ to $505.0\times$, while peak $R_0$ varies much less, from $0.744$ to $1.806$. This narrower range nonetheless spans the critical value $R_0=1$, so groups can fall into qualitatively different regimes, with network amplification either sustaining engagement or letting it die out. The two measures do not track each other, which is the central evidence for our main finding that exogenous response and endogenous spreading potential behave as distinct mechanisms in our estimates. Several comparisons illustrate this from different directions. AI Activists (pro) combines the highest peak $R_0$ in the study with a mid-ranking exogenous ratio, whereas Economics pairs the second-largest ratio with a near-critical $R_0$. Computer Science and Genetics/Biotech have nearly identical exogenous ratios but peak $R_0$ values that differ by a factor of two, and AI Research records the lowest exogenous ratio of any reliably fit group, yet still reaches an $R_0$ comfortably above $1$. 

Because $R_0(w)$ factors into network structure $\lambda_{\max}(A)$ and effective spreading rate $\beta_w/\gamma$ evaluated at the peak engagement window, the matched comparisons can be decomposed. The pro/anti gap reflects both: the pro network has about $1.3$ times the spectral radius ($49.3$ vs.\ $37.9$), and its effective spreading rate is about $1.3$ times higher. The Computer Science vs. Genetics/Biotech contrast is driven by effective spreading rate, against the structural difference: Genetics/Biotech has more than twice the spectral radius ($239.1$ vs.\ $108.8$) but roughly $4.3$ times lower effective spreading rate, so the model attributes its subcritical engagement to how little activity passes along its ties rather than to a sparse network. More generally, $\lambda_{\max}(A)$ varies roughly tenfold across reliably fit groups while peak $R_0$ varies only about $2.4$-fold, and the two groups with the largest spectral radii, Public Health/Epidemiology and Economics, have below-median peak $R_0$. Hence, network structure alone does not determine a group's spreading potential.

Topical proximity to the event also shapes when and whether engagement appears. DeepSeek R1's two-stage public reaction produces a small precursor rise shortly after the January 20 technical release in the groups closest to AI practice (AI Research, AI General Topic, Computer Science, and Engineering), while groups farther from the topic remain largely flat until the January 27 mainstream spike. Proximity also appears to govern which groups have enough repeat engagement to be fit reliably: the flagged groups are not the smallest in size (Table~\ref{tab:network_characteristics}), as Non-AI Artists ($N=2{,}398$) is the largest AI-related community and Clinical/General Medicine ($N=2{,}973$) exceeds several reliably fit disciplines. Network size alone thus does not predict whether a community's engagement takes the form of the sustained, repeated posting our model detects.

The mechanism-level measures are also distinct from overall engagement volume. AI General Topic and AI Activists (pro) reach nearly the same engagement rate ($21.1\%$ and $19.8\%$), yet AI General Topic has roughly half the exogenous ratio ($52.0\times$ vs.\ $111.7\times$) and a lower peak $R_0$ ($1.427$ vs.\ $1.806$). Aggregate volume alone thus cannot reveal which mechanism produced a response. Despite the variation in mechanism strengths, the timing of peak engagement relative to the event is comparatively consistent: excluding AI Activists (pro), lag to peak clusters tightly between $+1.3$ and $+1.6$ days across all other reliably fit groups, suggesting that the delay between the event and peak engagement varies far less than the strength of each mechanism.

\section{Discussion and Conclusion}
\label{sec:conclusion}

This paper developed a network-based SIR-type model that includes both exogenous engagement and network-driven spreading. The recovery rate, $\gamma$, and exogenous engagement rate, $\mu$, are estimated directly from the data, and the network spreading rate, $\beta_w$, is the only fitted parameter. We apply the model to the DeepSeek R1 release using two ways of grouping users: AI-related communities and academic disciplines. This allows us to compare how strongly each group responds directly to the event (the event-to-ambient exogenous ratio) with how strongly engagement spreads through the network (the peak endogenous spreading potential, $R_0(w)$). Our central finding is that these two quantities are not necessarily aligned: groups with a strong direct response to the event do not always show correspondingly strong network-driven spreading, and the reverse holds as well. This dissociation suggests that aggregate engagement volume alone can obscure meaningfully different underlying response mechanisms across communities. Decomposing $R_0(w)$ further shows that network structure alone does not determine a group's spreading potential: differences in effective spreading rate can offset structural differences or even reverse them, as when a denser network spreads less than a sparser one. Topical proximity to the event also influences both when engagement begins and whether it is sustained enough to be modeled reliably.

Several modeling choices bound these results. Starter-pack-based discovery skews toward already-prominent, well-followed accounts, so fitted parameters characterize engagement within a disproportionately prominent subnetwork rather than Bluesky's user base as a whole. Because the Recovered state is absorbing, the model cannot represent users who re-engage after a lapse and attributes that engagement to new participants. Fitted to the observed $I(t)$, it overestimates the number of distinct participants by $2.39\times$ for AI General Topic. We therefore do not compare groups by participant counts. This overcounting may also inflate $\beta_w$, and because we have not quantified whether it does so equally across groups, cross-group differences in $R_0$ should be read with this caveat. A degree-normalized spreading term motivated by \cite{feng2015competing} reduces the overestimate only to $2.21\times$, suggesting that an SIRS-type structure allowing return to engagement may be needed. Exposure is modeled only through the static follow graph, with other sources, such as algorithmic feeds, absorbed into $\mu(t)$. Because $\mu(t)$ departs from its ambient level only within a single event window, exogenous engagement outside it, such as the precursor response to DeepSeek R1's January 20 release, is attributed to network-driven spreading; a multi-pulse $\mu(t)$ would separate these contributions. Constant $\beta_w$ within each sub-window also causes fitted curves to smooth over and slightly undershoot the sharpest peaks (Figs.~\ref{fig:ai_topic_fit} and~\ref{fig:cs_fit}). Finally, we report point estimates; bootstrapping over nodes would quantify uncertainty in $R_0(w)$.

More broadly, our framework measures not only whether and how strongly a community reacts to a shared event, but also through which mechanism, and how that balance varies across communities. Because $\gamma$, $\mu$, and the adaptive windowing procedure are all estimated from each group's own data, nothing in the procedure is specific to AI discourse; applying it to a new event requires only changing the on-topic keyword filter. As a preliminary test, we applied it to Engineering's response to the June 14, 2025 ``No Kings'' rally, a political event unrelated to AI: engagement concentrated almost entirely in one day, with a large exogenous ratio ($210\times$), while $R_0$ quickly fell to near zero ($0.001$), indicating a response driven mainly by the event itself. Whether this fast-spike signature characterizes single-day mobilization events more generally, and how communities influence one another, are natural next questions. During data collection, we also explored the labels that Bluesky's labeler services \cite{atproto2026api} apply to accounts. Labels may be self-assigned, applied by independent third-party labelers, or generated automatically; some accounts in our networks, for instance, carry an \texttt{ai-hater} label from an independent labeler. These labels offer an additional dimension of information and raise a natural question for future work: how pro- and anti-AI sentiment evolve, similarly or differently, across communities around shared events.

\section*{Acknowledgment}

We thank George Mohler for helpful discussions and suggestions during the development of this work. We also acknowledge support from NSF DMS under Grant No. 2042413 and AFOSR MURI under Grant No. FA9550-22-1-0380.

\bibliographystyle{IEEEtran}
\bibliography{references}

\end{document}